\documentclass[10pt, conference]{IEEEtran}

\usepackage{xspace}
\usepackage{balance}
\usepackage{graphicx}
\usepackage[hyphens]{url}
\usepackage{epstopdf}
\usepackage{microtype}
\usepackage{booktabs}
\usepackage[table,xcdraw]{xcolor}
\usepackage{caption}
\usepackage{subcaption}
\usepackage{amsmath}
\usepackage{enumitem}
\usepackage{amssymb}
\usepackage[numbers]{natbib}
\usepackage{tikz}
\usepackage{pifont}
\usepackage{graphics}
\usepackage{array}
\usepackage{algorithm,algorithmic,refcount}

\newcommand{\PreserveBackslash}[1]{\let\temp=\\#1\let\\=\temp}
\newcolumntype{M}[1]{>{\PreserveBackslash\centering}m{#1}}
\newcolumntype{C}[1]{>{\PreserveBackslash\centering}p{#1}}
\newcolumntype{R}[1]{>{\PreserveBackslash\raggedleft}p{#1}}
\newcolumntype{L}[1]{>{\PreserveBackslash\raggedright}p{#1}}
\newcolumntype{P}[1]{>{\centering\arraybackslash}p{#1}}

\newcommand{\fakepar}[1]{\smallbreak\noindent}
\newcommand{\boldpar}[1]{\smallbreak\noindent\textbf{#1.}}

\newcommand{\ieee}{\mbox{IEEE~802.15.4}\xspace}

\newcommand{\blefive}{\mbox{BT\,5}\xspace}

\newcommand{\minimal}{\mbox{6TiSCH minimal}\xspace}

\usepackage{manyfoot}
\DeclareNewFootnote{A}
\DeclareNewFootnote{B}

\let\footnoteR\footnoteB
\let\footnote\footnoteA
\iftrue
    \newcommand{\michael}[1]{\footnoteR{{\color{red}\bf MB: #1}\color{red}}}
    \newcommand{\israat}[1]{\footnoteR{{\color{red}\bf IH: #1}\color{red}}}
    \newcommand{\atis}[1]{\footnoteR{{\color{red}\bf AE: #1}\color{red}}}
\else
    \newcommand{\michael}[1]{}
    \newcommand{\israat}[1]{}
    \newcommand{\atis}[1]{}
\fi
\title{BlueTiSCH: A Multi-PHY Simulation of Low-Power 6TiSCH IoT Networks}

\iffalse
\author{
\IEEEauthorblockN{Anonymous Author(s)} 
}
\else
\author{
\IEEEauthorblockN{
Chloe Bae\IEEEauthorrefmark{1}, 
Shiwen Yang\IEEEauthorrefmark{1}, 
Michael Baddeley\IEEEauthorrefmark{2},
Atis Elsts \IEEEauthorrefmark{3},
and
Israat Haque\IEEEauthorrefmark{1}\\
} \vspace{-2.50mm}\\ 
\IEEEauthorblockA{\IEEEauthorrefmark{1}Department of Computer Science, Dalhousie University, Halifax, Canada -- \url{{chloe.bae; shiwen.yang; israat}@dal.ca} }
\IEEEauthorblockA{\IEEEauthorrefmark{3}Institute of Electronics and Computer Science (EDI), Riga, Latvia -- \url{atis.elsts@edi.lv} } 
\IEEEauthorblockA{\IEEEauthorrefmark{2}Technology Innovation Institute (TII), Abu Dhabi, UAE -- \url{michael.baddeley@tii.ae} } 
\vspace{-5.00mm}\\
}
\fi

\IEEEoverridecommandlockouts\IEEEpubid{\makebox[\columnwidth]{ 978-1-6654-3540-6/22~\copyright~2022 IEEE \hfill} \hspace{\columnsep}\makebox[\columnwidth]{ }}

\begin{document}
\maketitle
\begin{abstract}
Low-power wireless IoT networks have traditionally operated over a single physical layer (PHY) -- many based on the \ieee standard. However, recent low-power wireless chipsets offer both the \ieee and all four PHYs of the Bluetooth\,5 (\blefive) standard. This introduces the intriguing possibility that IoT solutions might not necessarily be bound by the limits of a single PHY, and could actively or proactively adapt their PHY depending on RF or networking conditions (e.g., to offer a higher throughput or a longer radio range). Several recent studies have explored such use-cases. However, these studies lack comprehensive evaluation over various metrics (such as reliability, latency, and energy) with regards to scalability and the Radio Frequency (RF) environment. In this work we evaluate the performance of \ieee and the four \blefive 2.4GHz PHY options for the recently completed IETF 6TiSCH low-power wireless standard. To the best of our knowledge, this is the first work to directly compare these PHYs in identical settings. Specifically, we use a recently released 6TiSCH simulator, TSCH-Sim, to compare these PHY options in networks of up to 250 nodes over different RF environments (home, industrial, and outdoor), and highlight from these results how different PHY options might be better suited to particular application use-cases.
\end{abstract}
\vspace{-2mm}

\section{Introduction}
\label{sec:introduction}

Internet of Things (IoT) communications have typically focused on a single low-power wireless standard. Whether \ieee, Bluetooth, or LoRa, the overall aim of these specifications has been to reduce device power consumption while maintaining connectivity and availability requirements for the desired use-case~\cite{IFIP,softiot}. \ieee, in particular, has served as the underlying physical (PHY) and medium access control (MAC) layer for a number of industrial networking standards, and recently completed efforts from the IETF 6TiSCH Working Group (WG)~\cite{ietf_6tisch} has introduced IPv6 enabled scheduling and networking mechanisms for the Time Slotted Channel Hopping (TSCH) MAC option introduced as part of the \ieee-2015 amendment.

While TSCH enhances the reliability of IoT wireless network communications by devising communication \texttt{SlotFrames} across time and frequency, 6TiSCH provides mechanisms for efficiently scheduling those communications between nodes. Co-located nodes are able to concurrently transmit on orthogonal channels at each timeslot due to the available channel diversity \cite{ietf_6tisch}; thus, improving the capacity of the network over traditional Carrier Sense Multiple Access (CSMA) approaches. However, the 6TiSCH standard is, in fact, agnostic to the underlying PHY layer though the majority of literature surrounding 6TiSCH has almost exclusively focused on \ieee. With modern low-power wireless chipsets (such as the \texttt{Nordic~nRF52840}~\cite{NordicSemi}, and \texttt{TI~CC2650}~\cite{TexasInstru}) supporting multiple PHY standards on a \textit{single radio}, 6TiSCH opens the intriguing possibility of supporting PHYs other than \ieee. Specifically, the nRF52840 supports \ieee, as well as all four \blefive PHY options. This multi PHY support can allow networks to operate over the \blefive\,125K option for long-range applications. In contrast, applications that require greater throughput over shorter distances can operate over \blefive\,2M high data-rate PHY. 

There have been several recent studies proposing a multi-PHY approach to 6TiSCH IoT networks~\cite{parent-selection, 6pp, NoFreeLunch, g6TiSCH}. These works mostly focused on a limited number of nodes (up to 100), a single environment (indoors or outdoors), or a single scheduler (\minimal or Orchestra). Thus, we perform an extensive performance evaluation of five PHYs over: \emph{(i)} both the \minimal and Orchestra schedulers, \emph{(ii)} networks of up to 250 nodes, \emph{(iii)} across three different RF environments (\textit{home}, \textit{industrial}, and \textit{outdoor}). Specifically, we implement and evaluate \ieee and all four \blefive PHYs over 6TiSCH network simulator, TSCH-Sim~\cite{TSCHsim}. Thus, users can choose an appropriate combination of PHY and scheduler for their targeted environment for a better reliability, latency, and energy usage. Furthermore, we make available our simulation code and data to the community for reproducibility and extension\footnote{https://github.com/mbaddeley/tsch-sim-mphy}\footnote{https://github.com/dohibae/tsch-project}. The evaluation results reveal that  BLE~500K is the best option for applications that require  high  PDR and the best replacement for \ieee if used as a standalone PHY. At the same time, the  uncoded BLE~1M  and BLE~2M options achieve lower latency and lower energy usage.

This paper is structured as follows. In Section~\ref{sec:background}, we provide an overview of the RPL routing layer (Layer 3), the \minimal and Orchestra Schedulers (Layer 2), and the multiple PHY standards (Layer 1) -- explaining how the different aspects of each PHY may impact the higher layers. In Section~\ref{sec:related-work}, we summarize recent literature examining multi-PHY 6TiSCH approaches. Section~\ref{sec:evaluation} provides details of our simulation setup following extensive performance results of each PHY. Finally, Section~\ref{sec:conclusions} concludes our work.
\section{Background}
\label{sec:background}

We provide an overview of 6TiSCH protocol, RPL Routing (Layer 3), 6TiSCH Schedulers (Layer 2) and the standards of multiple PHY’s (Layer 1), relevant to the study.

\textbf{6TiSCH.}
IETF 6TiSCH, the architecture for IPv6 over the TSCH mode of \ieee, provides mechanisms for efficient scheduling and coordination of the TSCH slot frame. \ieee-2015 did not address \emph{how} communication can be scheduled while the standard introduced TSCH as a means for synchronizing and scheduling co-located communications over orthogonal channels. 6TiSCH fills this gap by allowing schedulers a way to manage communications across a network to avoid self-interference, i.e., contention between wireless devices, and provide optimized schedules for packet transmissions, reception, and sleeping (idling) for each time slot. Specifically, 6TiSCH has shared cells (shared with neighboring nodes) and dedicated cells (of which task is specific and fixed for a certain time slot). Time synchronization of 6TiSCH cells is achieved by using the information contained in the Enhanced Beacons (EBs) and in the Keep Alive (KA) packets, which are periodically broadcast via the shared 6TiSCH cells. 

\boldpar{RPL Routing Algorithm (Layer 3)}
Low-power wireless networks typically employ the lightweight distance-vector-based RPL routing protocol~\cite{ietf_rpl}. RPL uses a tree-like graph called a DODAG (Direction-Oriented Directed Acyclic Graph), ideal for data-collection use-cases. Usually, 6TiSCH is not dependent on RPL, but the two are often used in conjunction, e.g., in the Orchestra scheduler~\cite{duquennoy2015orchestra}. 
Specifically, when employing RPL's Non-Storing mode, RPL control traffic can create overhead as the network scales~\cite{baddeley2020thesis}. We can mitigate this overhead by using orthogonal wireless communication channels for the control packets separated from the channels used for data packets while using 6TiSCH.


\boldpar{Scheduling Algorithm (Layer 2)}
Within 6TiSCH, schedulers play an important role on network performance, determining how the 6TiSCH cells coordinate with each other. For instance, Orchestra~\cite{duquennoy2015orchestra} allows nodes to autonomously manage their own schedules, assigning roles (e.g., advertising, dedicated, or shared) and particular tasks (e.g., sleep, transmit a packet, or receive a packet) based on current traffic. On the other hand, 6TiSCH Minimal Scheduling Function~(MSF)~\cite{rfc9033} is a minimal scheduler implementation analogous to slotted CSMA. This study examines PHY performance over these two schedulers.

\boldpar{Physical Layer Standards (Layer 1)}
Traditionally both 6TiSCH and RPL routing protocol have been practically based on \ieee PHY, with much of the existing literature focused on the 2.4\,GHz OQPSK-DSSS variant. While this allows a greater data-rate (250\,kbps) than the sub-GHz \ieee PHYs (typically a few tens of kbps), this severely limits radio range. Recently introduced multi-PHY chips (in particular, the nRF52840~\cite{NordicSemi} has become a popular platform in the low-power wireless community) target \emph{both} \ieee \emph{and} \blefive (Bluetooth Low Energy) PHY configurations - introducing the possibility of higher data rates (\blefive\,2M and \blefive\,1M) and longer radio ranges (\blefive\,500K and \blefive\,125K). This study examines 6TiSCH performance over \ieee OQPSK-DSSS and all four \blefive PHY options for both Orchestra and MSF, taking into account typical radio sensitivity and ranges in \textit{home}, \textit{industrial}, and \textit{outdoor} environments (see Table~\ref{table:sim_settings}).


\begin{table}
\vspace{+2mm}
\caption{Simulation Settings.}
\label{table:sim_settings}
\renewcommand{\arraystretch}{1.2}
\centering
\footnotesize
 \begin{tabular}{ m{10em} m{2.2em} | m{2.2em} | m{2.2em} |  m{2.2em} | m{3em} }
 \toprule
 \bfseries Parameter & \bfseries \blefive 2M & \bfseries \blefive 1M & \bfseries \blefive 500K  & \bfseries \blefive 125K & \bfseries IEEE  802.15.4 \\
 \midrule
    \multicolumn{4}{l}{\textbf{MAC layer (TSCH) settings}}\\
        Scheduler & \multicolumn{5}{c}{Orchestra / \minimal} \\
        ACK size, bytes & \multicolumn{4}{c|}{2} & 17 \\ 
        ACK wait time, µs & \multicolumn{4}{c|}{150} & 400 \\ 
        RX wait time, µs & \multicolumn{4}{c|}{150} & 2,200 \\ 
        MAC header size, bytes & \multicolumn{4}{c|}{6} & 23\\
        Slot duration, µs & 1,064 & 2,120 & 4,542 & 17,040 & 4,256 \\ 
\midrule
    \textbf{Other layer settings} \\
        Routing Protocol  & \multicolumn{5}{c}{RPL} \\
        Traffic pattern & \multicolumn{5}{c}{Peer-to-peer} \\ 
        IP fragmentation & \multicolumn{5}{c}{No} \\ 
        App packet period, sec & \multicolumn{5}{c}{160} \\
        App packet size, bytes & \multicolumn{4}{c|}{251} & 102\\
        PHY overhead (bytes) & 9 & 8 & 26.875 & 9 & 8 \\ 
        Byte duration, µs & 4 & 8 & 16 & 64 & 32 \\ 
\midrule
    \textbf{RF env. settings~\cite{NistPap02}} \\
        Radio Medium & \multicolumn{5}{c}{Unit Disk Graph (UDGM)} \\
        Co-channel rejection, dB & \multicolumn{4}{c|}{-8} & -3 \\
        Home\,@\,0\,dBm & 23\,m & 26\,m & 40\,m & 43\,m & 30.5\,m \\
        Industrial\,@\,10\,dBm & 73.6\,m & 92\,m & 184\,m & 368\,m & 175\,m \\
        Outdoor\,@\,10\,dBm & 170\,m & 212\,m & 413\,m & 473\,m & 346\,m \\
\bottomrule
\end{tabular}
\vspace{-3.00mm}
\end{table}

\begin{figure*}[t!]
	\centering
	\begin{subfigure}[t]{.65\columnwidth}
		\centering
		\includegraphics[width=1.\columnwidth]{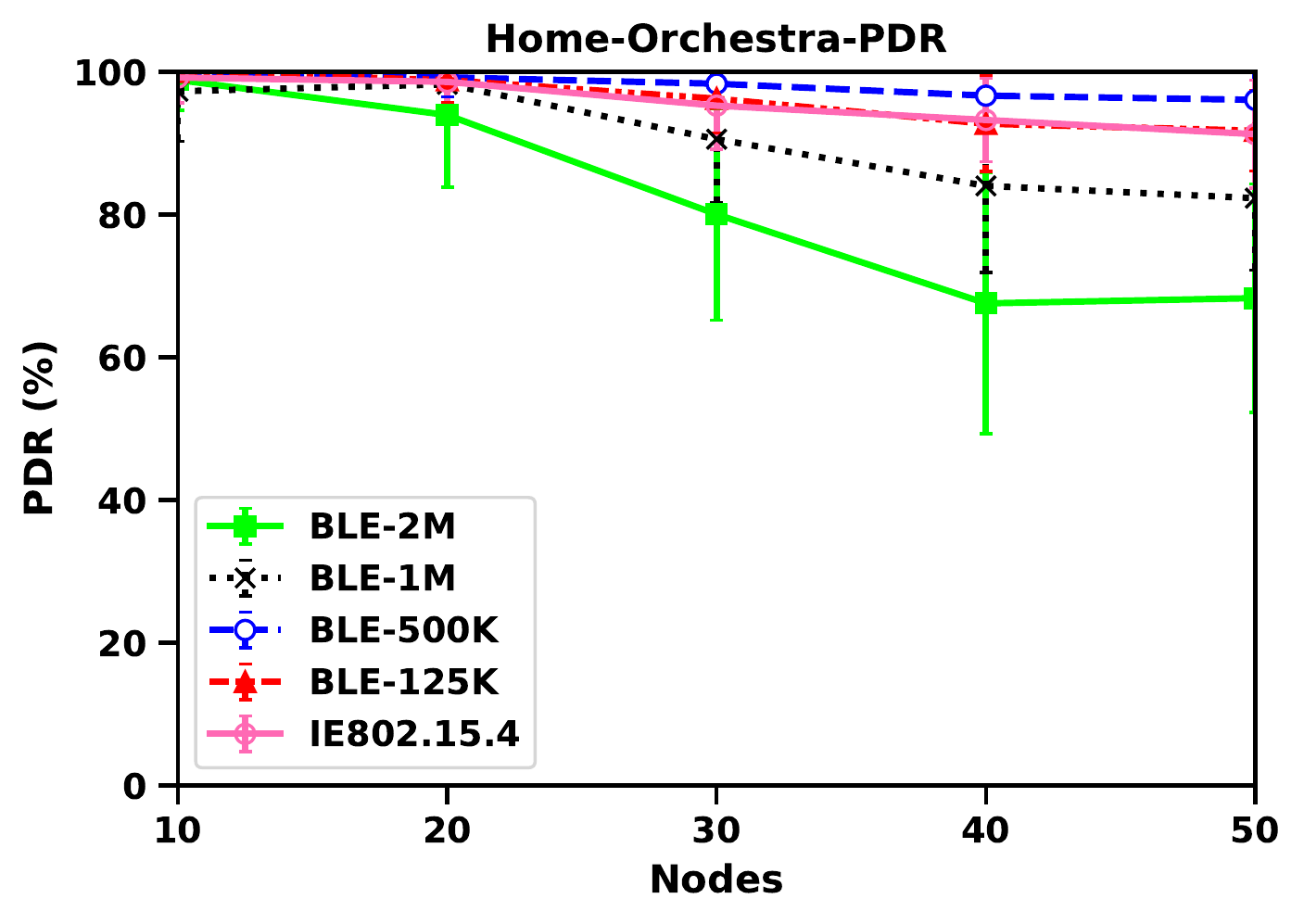}
		\caption{Home using Orchestra}
		\label{fig:home-orch-pdr}
	\end{subfigure}%
	\begin{subfigure}[t]{.65\columnwidth}
		\centering
		\includegraphics[width=1.\columnwidth]{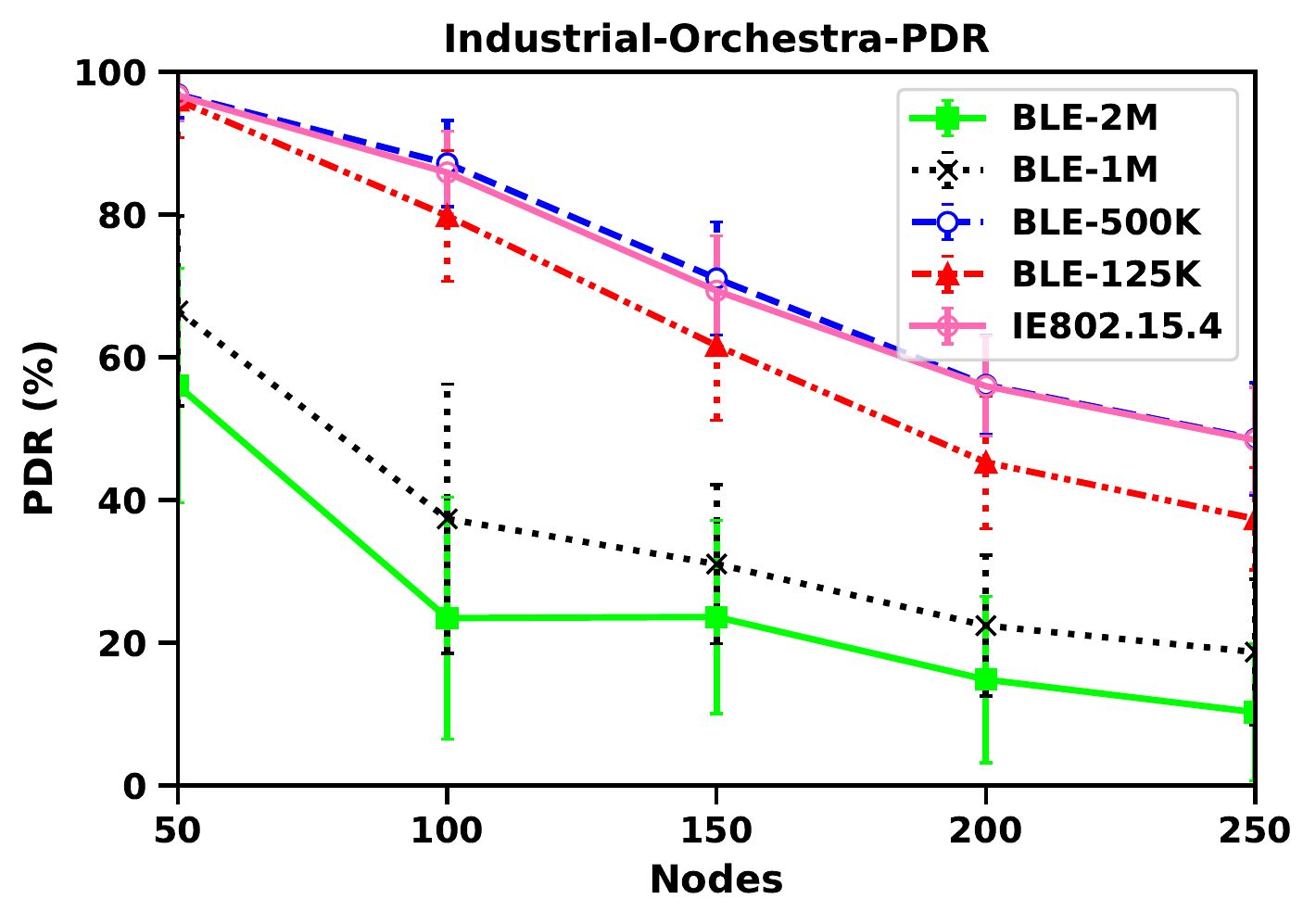}
		\caption{Industrial using Orchestra}
		\label{fig:indu-orch-pdr}
	\end{subfigure}%
	\begin{subfigure}[t]{.65\columnwidth}
		\centering
		\includegraphics[width=1.\columnwidth]{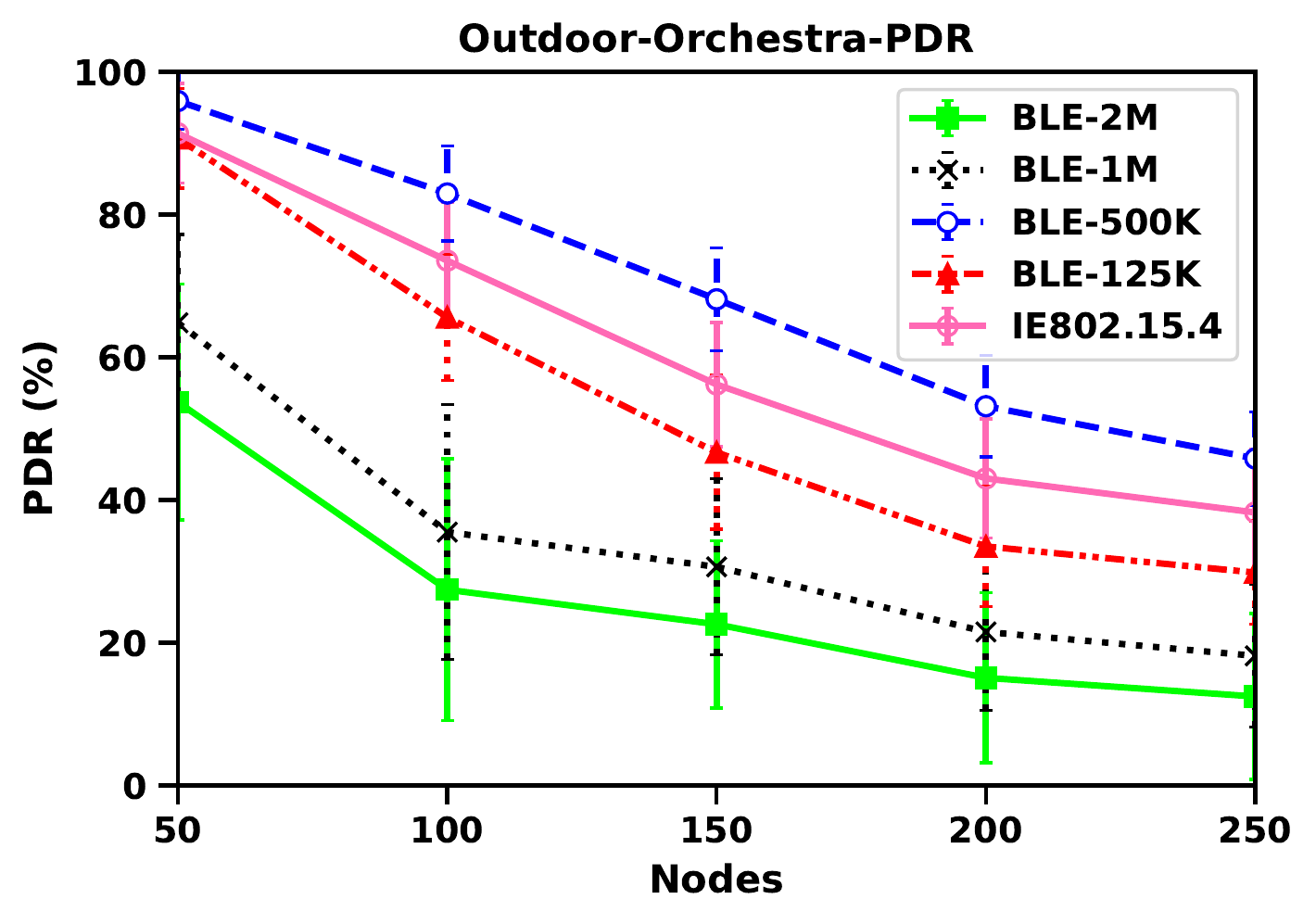}
		\caption{Outdoor using Orchestra}
		\label{fig:out-orch-pdr}
	\end{subfigure}%
	\hfill
	\begin{subfigure}[t]{.65\columnwidth}
		\centering
		\includegraphics[width=1.\columnwidth]{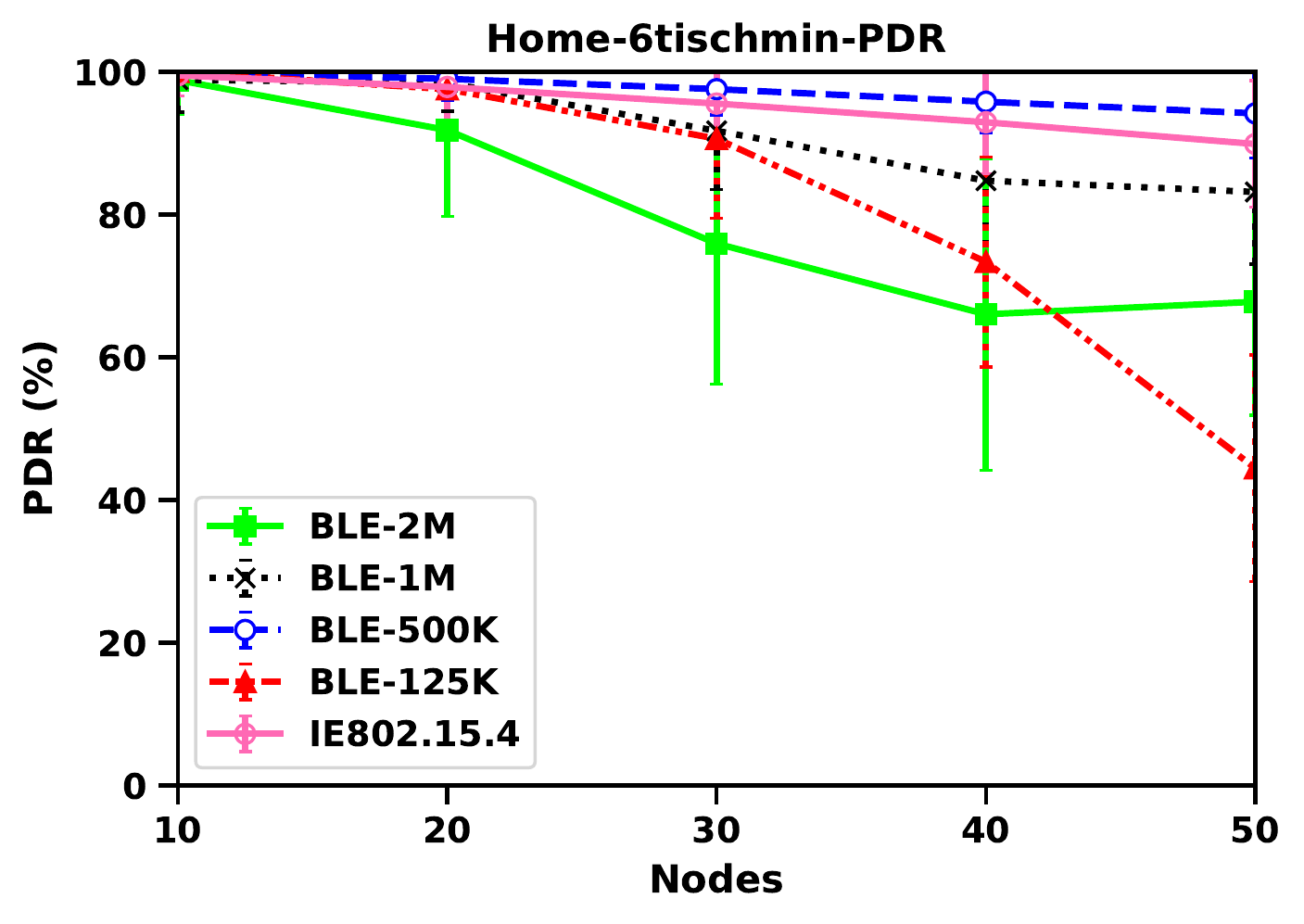}
		\caption{Home using \minimal}
		\label{fig:home-6ti-pdr}
	\end{subfigure}
	\begin{subfigure}[t]{.65\columnwidth}
		\centering
		\includegraphics[width=1.\columnwidth]{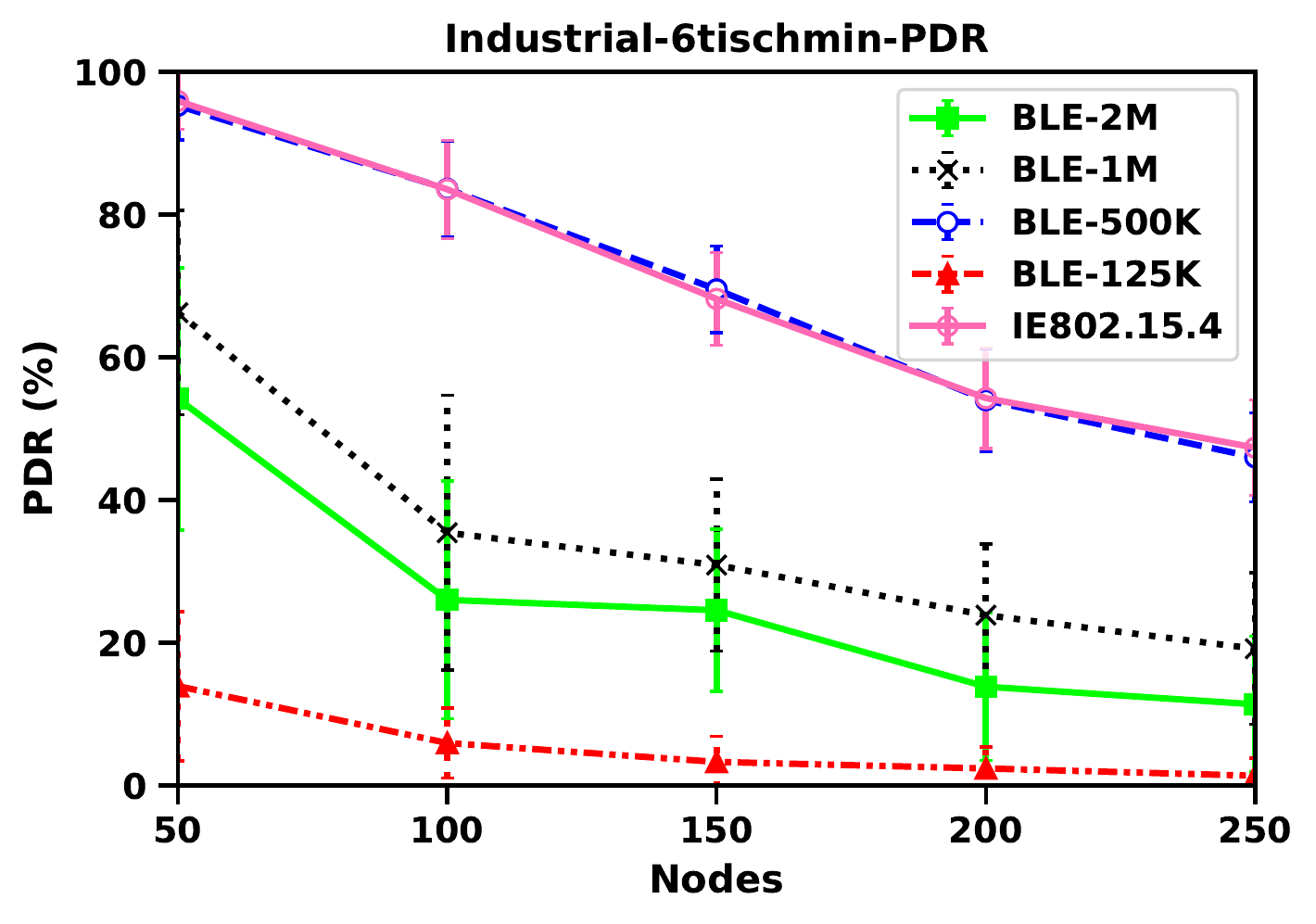}
		\caption{Industrial using \minimal}
		\label{fig:indu-6ti-pdr}
	\end{subfigure}
	\begin{subfigure}[t]{.65\columnwidth}
		\centering
		\includegraphics[width=1.\columnwidth]{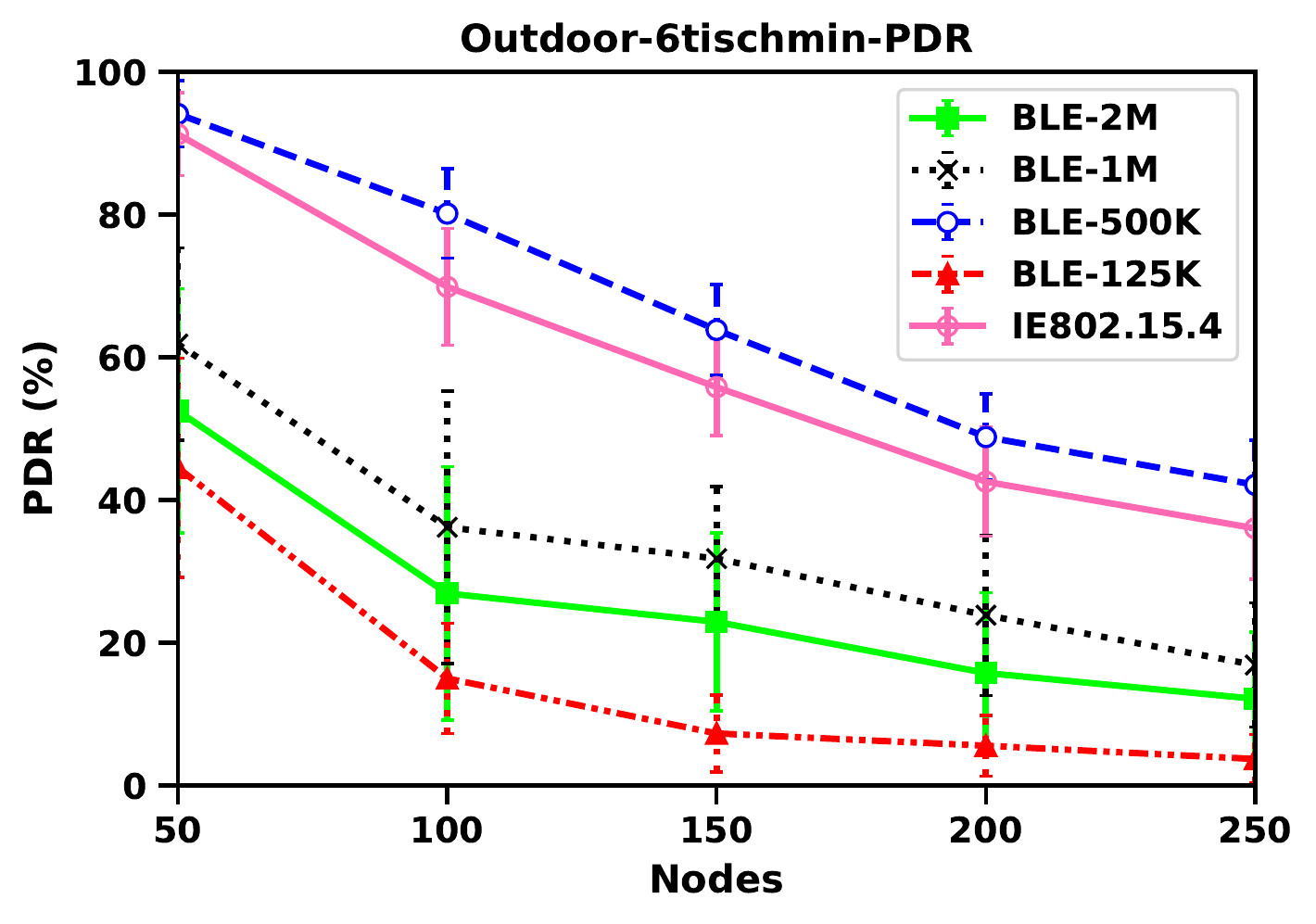}
		\caption{Outdoor using \minimal}
		\label{fig:out-6ti-pdr}
	\end{subfigure}
	\vspace{-0.50mm}
	\caption{Packet Delivery Ratio (\%).}
	\vspace{-2.0mm}
	\label{fig:pdr}
\end{figure*}

\section{Related work} 
\label{sec:related-work}

The authors of~\cite{NoFreeLunch,g6TiSCH} use differently modulated physical layers (FSK (low-rate), OFDM (high-rate), and O-QPSK (medium-rate)) to achieve runtime PHY selection in Industrial IoT (IIoT). They show that the long-ranged, low-rate PHY exhibits the best reliability and latency at the cost of significant (10x) degrade in battery life compared to O-QPSK, which exhibits the opposite trend. OFDM offers a balance between FSK and O-QPSK. Distributed PHY with a parent selection scheme is proposed in \cite{parent-selection}. Although all the above works performed practical experiments, they are confined to a limited number of nodes (at the scale of 50). IIoT networks can have various nodes ranging from low (below 50) to high (above 100) in different indoors and outdoors.    

\citet{badihi2019system} develop a system-level simulator to evaluate \blefive PHYs at scale. Specifically throughput, packet error rate (PER), end-to-end delay, and battery life. \blefive\,2M offers the highest throughput, lowest PER, lowest delay, and longest battery life due to the high bit rate and fewer collisions. Conversely, the \blefive\,125K coded PHY has the worst performance due to its large packet size and collisions.

\citet{schedulers} move one layer above from PHY and focuses on the performance of the two schedulers: Orchestra and MSF, over varying network density and packet rates at a scale of 98 nodes. The simulation results reveal that Orchestra is more reliable due to fewer collisions and prioritizing between slot frames and consumes less energy due to more sleep time slots. On the contrary, MSF achieves a better latency due to the low number of time slots and consumes less computing/memory resources. However, the authors in both papers consider only one use case (e.g., open office space) with at most 100 nodes that do not cover various indoor and outdoor IIoT environments and scales. \cite{6pp}~and~\cite{OR} both combine TSCH with Synchronous Flooding-based communications for high reliability signalling. Again, the solution is tested over a limited number of nodes in an indoor environment.

\section{Evaluation Setup and Metrics} \label{sec:evaluation}
We implement and evaluate the five PHYs (\blefive and \ieee), and two 6TiSCH schedulers (Orchestra and MSF) in the TSCH-Sim simulator \cite{TSCHsim}.
TSCH-Sim is a recent protocol-level simulator for investigating TSCH networks at scale and provides a 6TiSCH stack implemented on top of \ieee PHY. We modified TSCH-Sim parameters and the original simulation code to take into account the differences between the \ieee and \blefive standards, including different data rates and radio ranges (Table~\ref{table:sim_settings}), where radio ranges are estimated based on~\cite{NistPap02}.

Network topology is a randomly generated mesh without any disconnected partitions, while mesh density (i.e., degree per node) depends on the PHY range. 
For the shortest-range PHY, \blefive\,2M, the average node degree is set to 6.5 when generating the topology.
The same node locations are reused for other PHY layers, meaning that the number of connections per node is higher in these longer-range PHYs.
We then varied the number of nodes depending on the network type: for the home environment, a relatively small number of nodes are simulated, ranging between 10, 20, 30, 40, and 50 nodes. We considered 50, 100, 150, 200, and 250 nodes for industrial and outdoor environments to better represent the scale of such deployments (Table~\ref{table:sim_settings}). 

We repeated each simulation 100 times to calculate the mean and standard deviation, and each individual simulation was run for 300 seconds. Specifically, we evaluate across the following metrics:
\boldpar{Packet Delivery Ratio (PDR)} The ratio of the total number of packets delivered to the destination nodes to the total number of packets sent from the source nodes, and measures communication reliability. 

\boldpar{Latency} Time interval between the generation and reception of an application-layer packet. Both physical-layer (e.g., data rate) and higher-layer parameters (e.g., number of MAC retransmissions, routing topology) have impact on latency. 

\begin{figure*}[t]
	\vspace{+1.75mm}
	\centering
	\begin{subfigure}[t]{.65\columnwidth}
		\centering
		\includegraphics[width=1.\columnwidth]{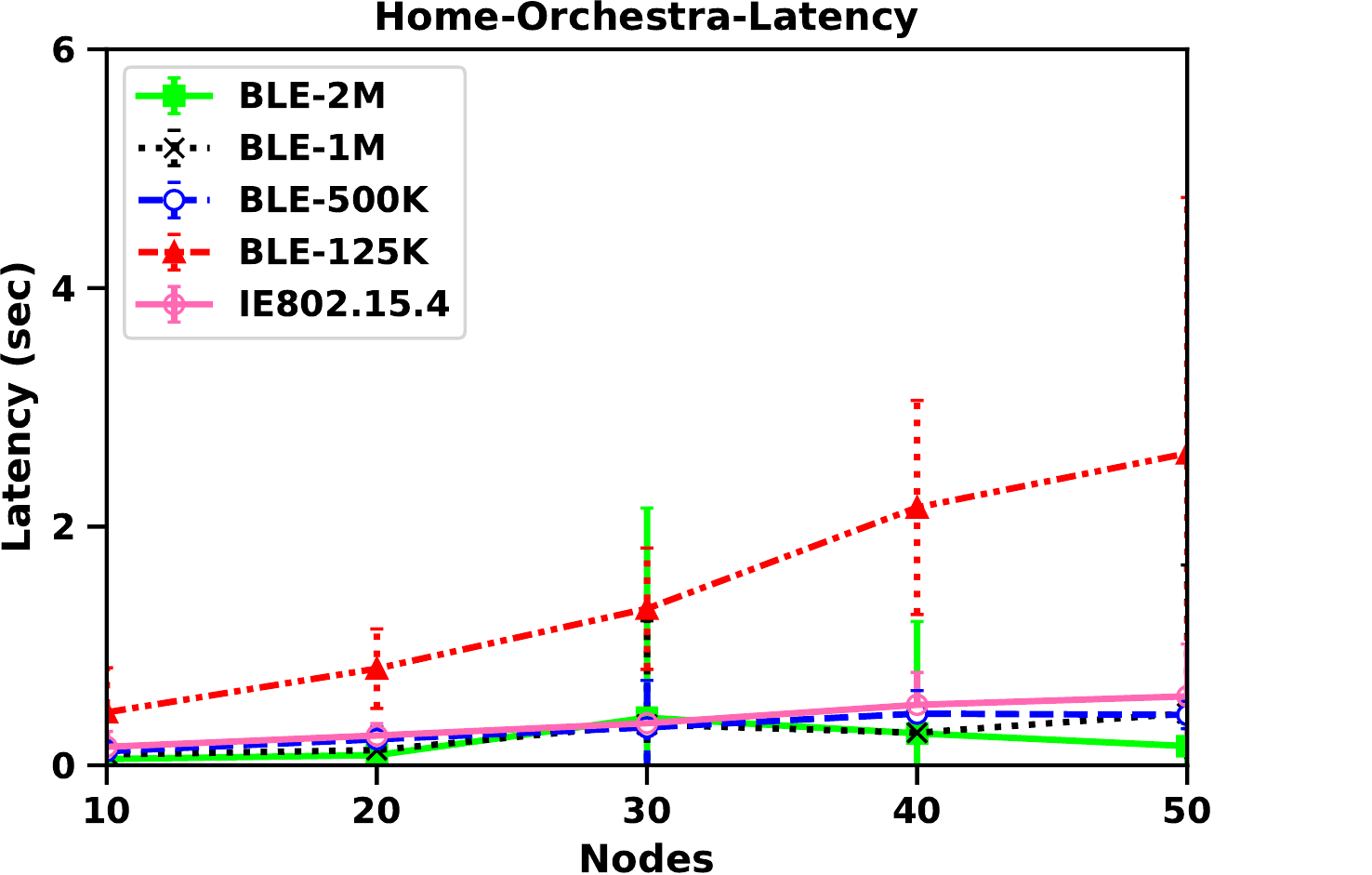}
		\caption{Home using Orchestra}
		\label{fig:home-orch-lat}
	\end{subfigure}%
	\begin{subfigure}[t]{.65\columnwidth}
		\centering
		\includegraphics[width=1.\columnwidth]{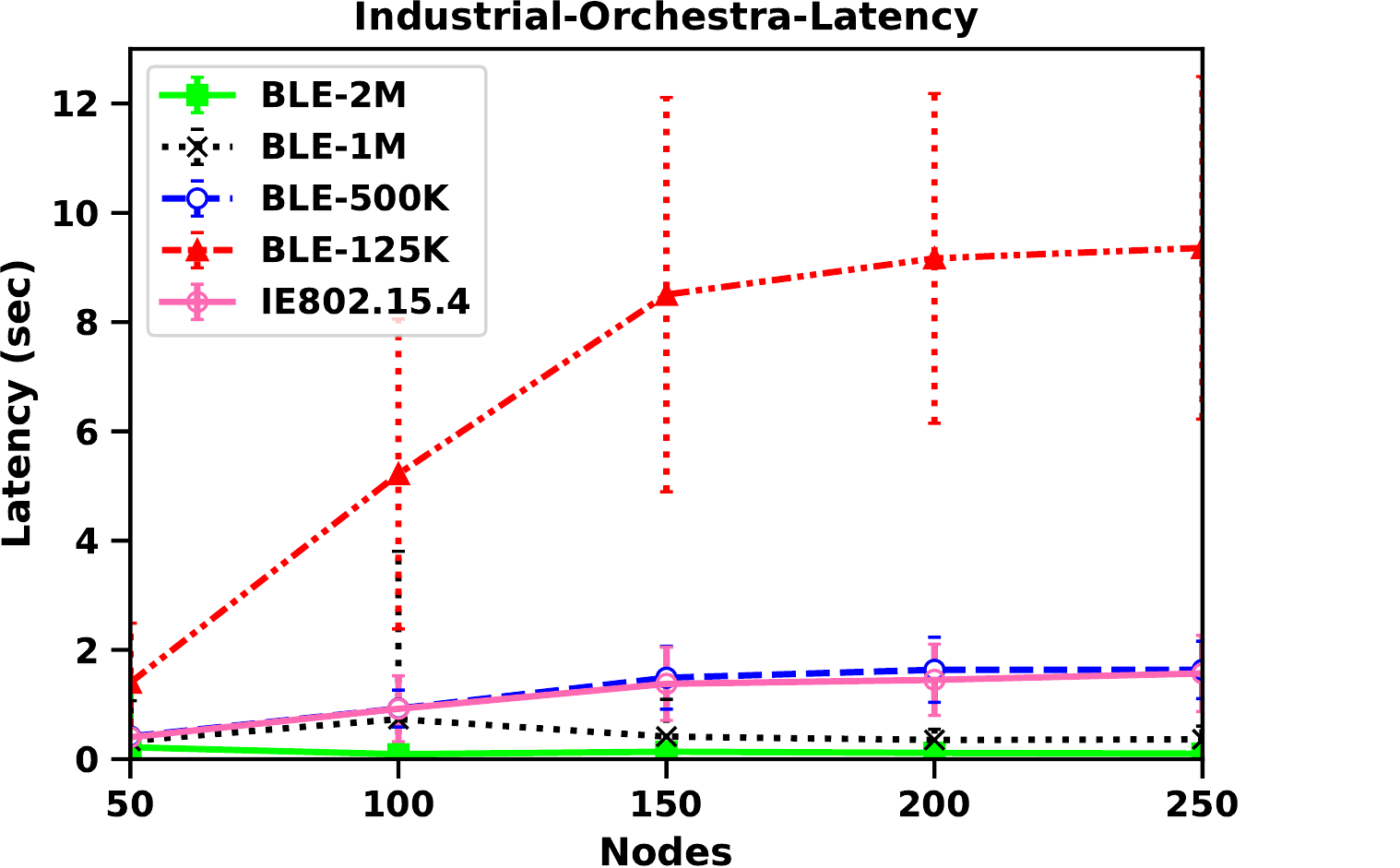}
		\caption{Industrial using Orchestra}
		\label{fig:indu-orch-lat}
	\end{subfigure}%
	\begin{subfigure}[t]{.65\columnwidth}
		\centering
		\includegraphics[width=1.\columnwidth]{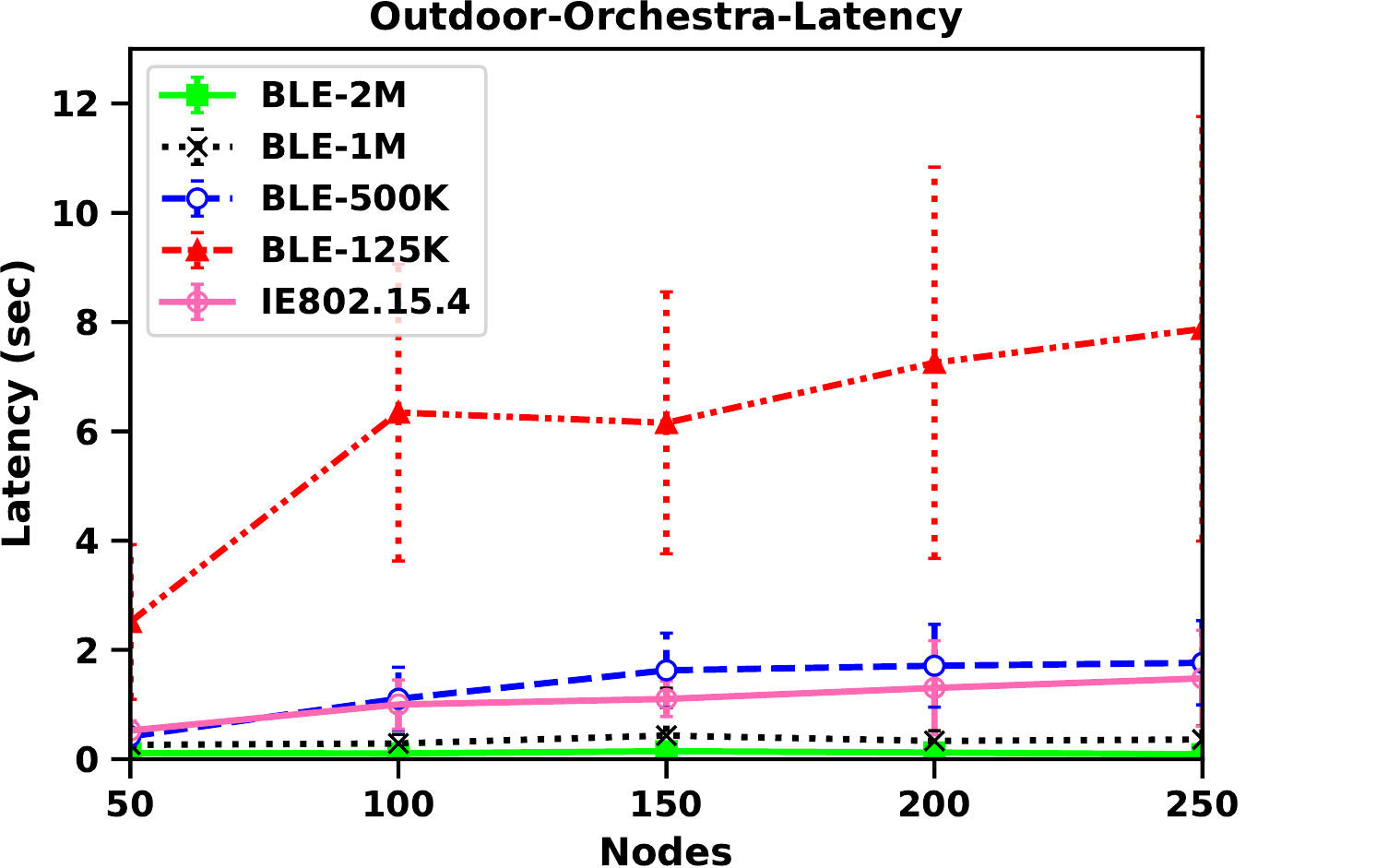}
		\caption{Outdoor using Orchestra}
		\label{fig:out-orch-lat}
	\end{subfigure}%
	\hfill
	\begin{subfigure}[t]{.65\columnwidth}
		\centering
		\includegraphics[width=1.\columnwidth]{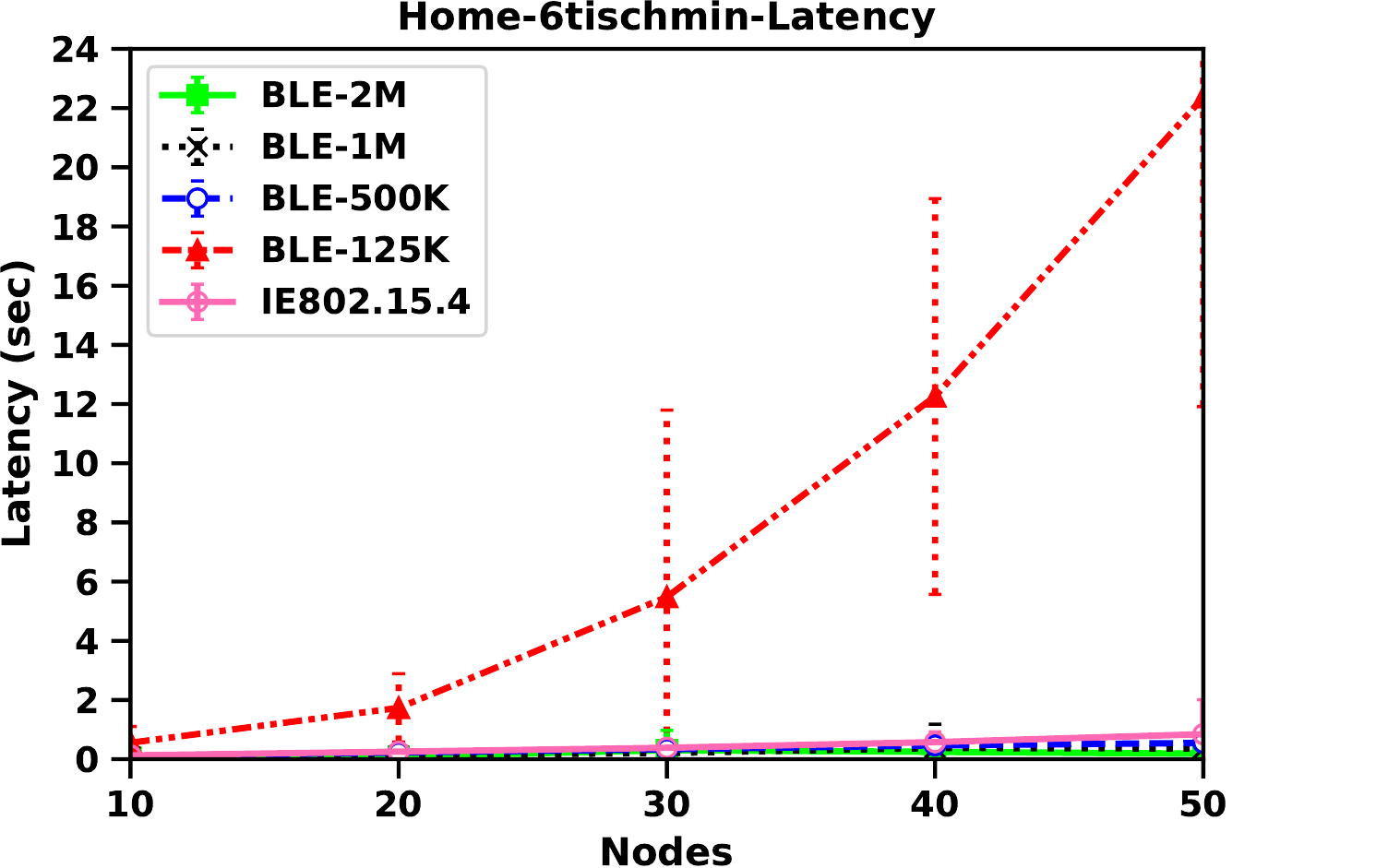}
		\caption{Home using \minimal}
		\label{fig:home-6ti-lat}
	\end{subfigure}
	\begin{subfigure}[t]{.65\columnwidth}
		\centering
		\includegraphics[width=1.\columnwidth]{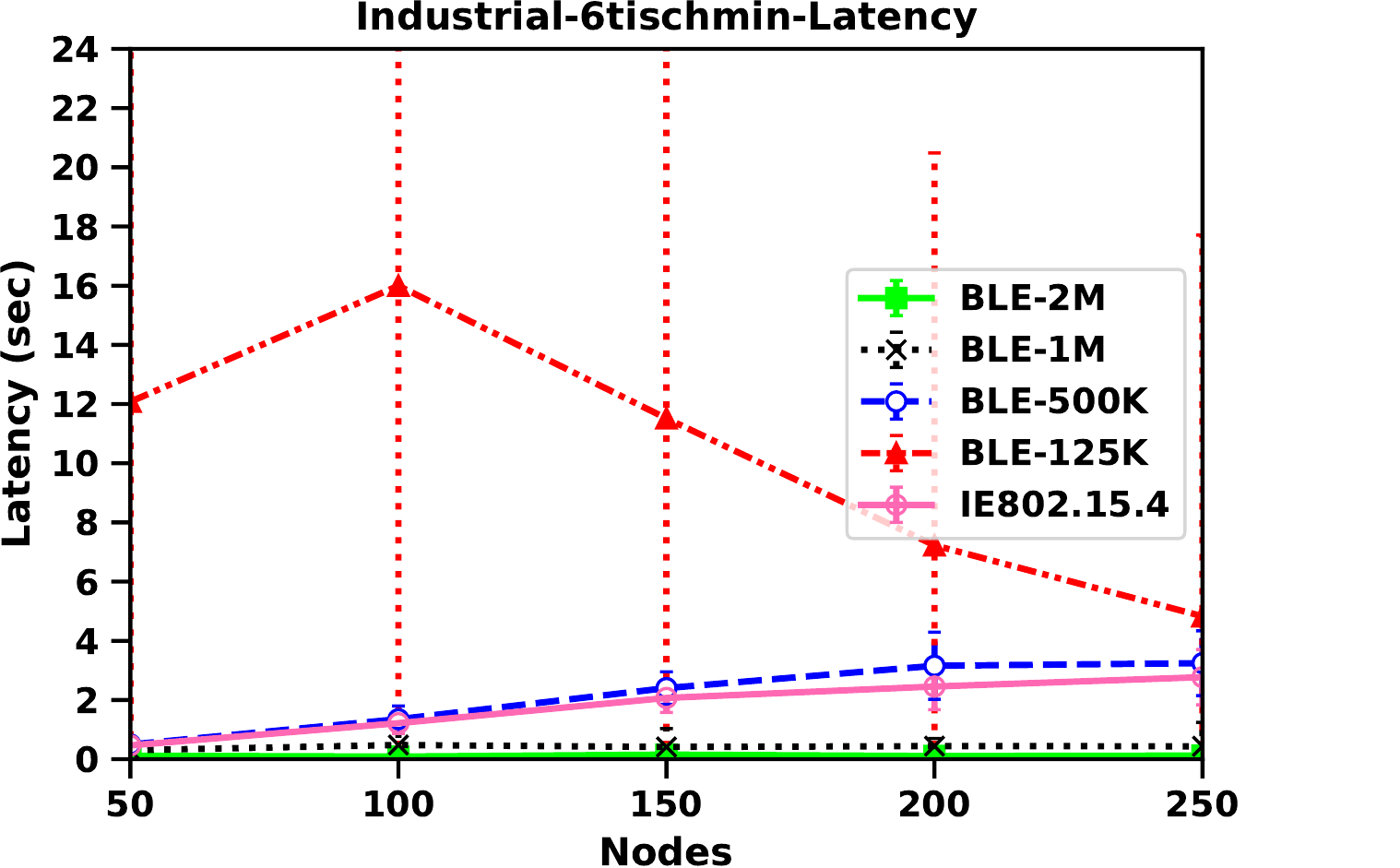}
		\caption{Industrial using \minimal}
		\label{fig:indu-6ti-lat}
	\end{subfigure}
	\begin{subfigure}[t]{.65\columnwidth}
		\centering
		\includegraphics[width=1.\columnwidth]{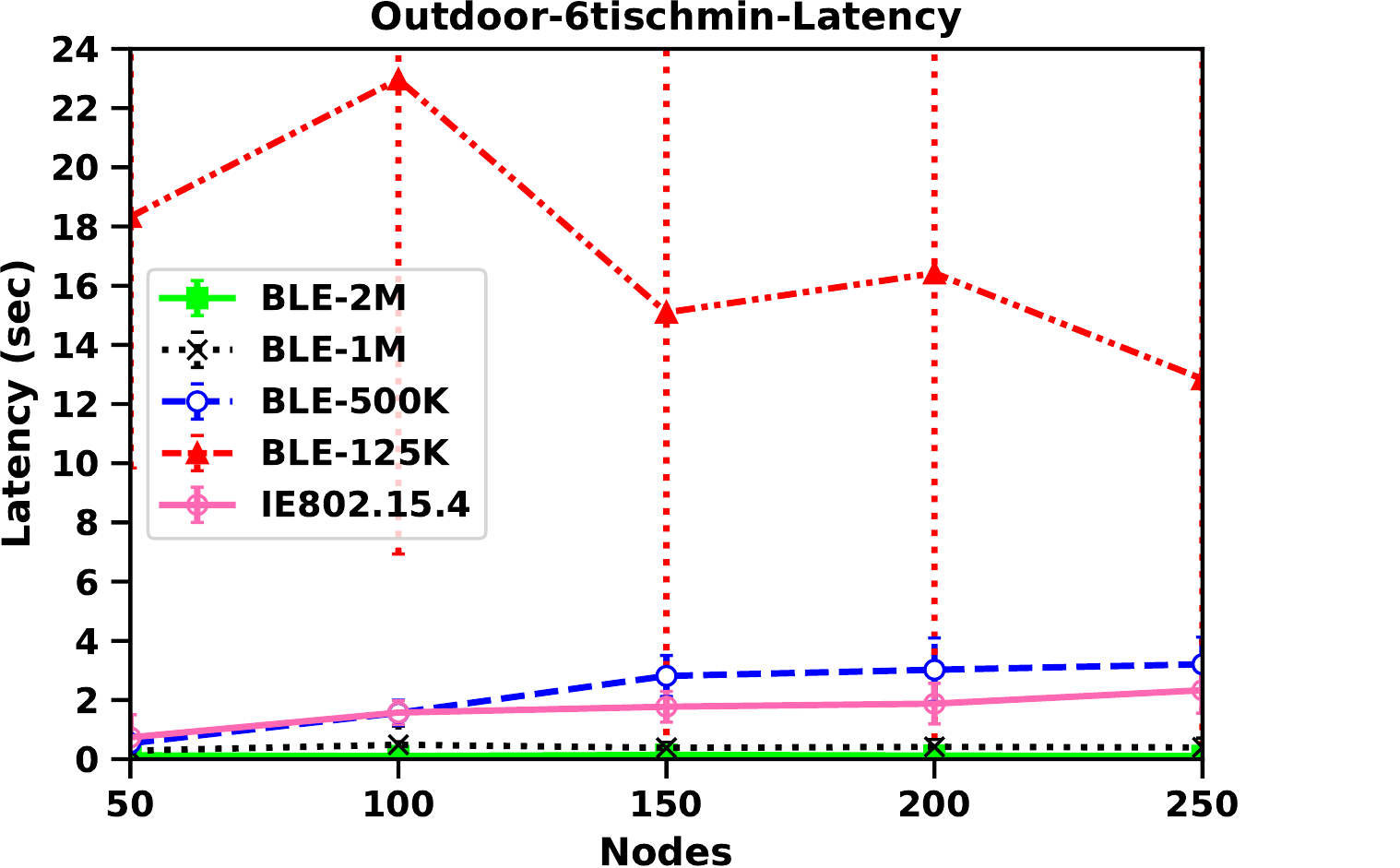}
		\caption{Outdoor using \minimal}
		\label{fig:out-6ti-lat}
	\end{subfigure}
	\vspace{-0.50mm}
	\caption{End-to-end latency in seconds.}
	\vspace{-2.0mm}
	\label{fig:latency}
\end{figure*}

\boldpar{Radio Duty Cycle (RDC)} The ratio between the time the radio is on and the total simulation time. The RDC we report only accounts nodes that have successfully joined the TSCH network and ignores the time spent in the scanning phase, as we are not focused on investigating the network formation performance in this paper. RDC strongly correlates with the total energy usage. 

\begin{figure*}[t]
	\vspace{+10mm}
	\centering
	\begin{subfigure}[t]{.65\columnwidth}
	    \vspace{+1mm}
		\centering
		\includegraphics[width=1.\columnwidth]{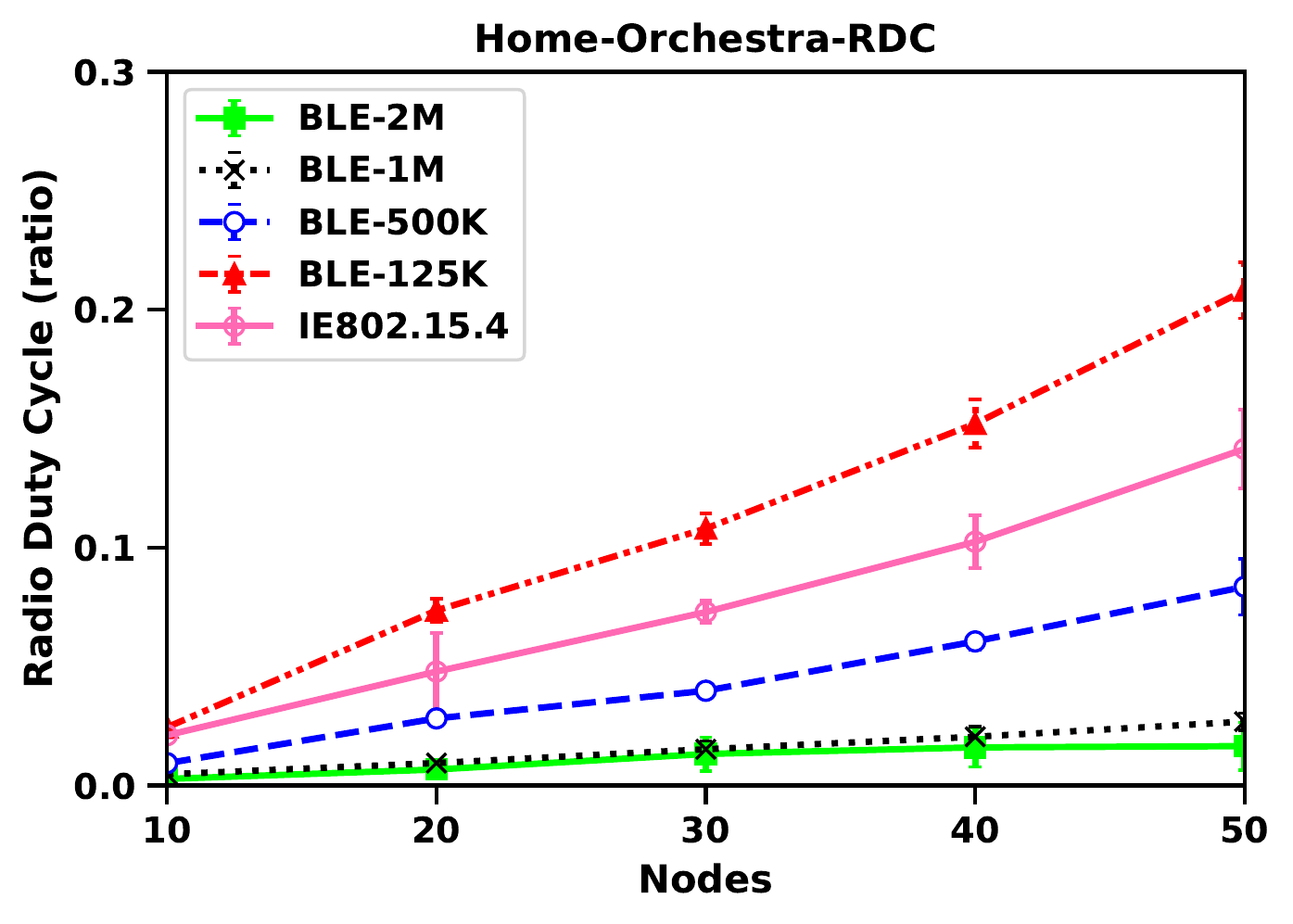}
		\caption{Home using Orchestra}
		\label{fig:home-orch-dc}
	\end{subfigure}%
	\begin{subfigure}[t]{.65\columnwidth}
	    \vspace{+1mm}
		\centering
		\includegraphics[width=1.\columnwidth]{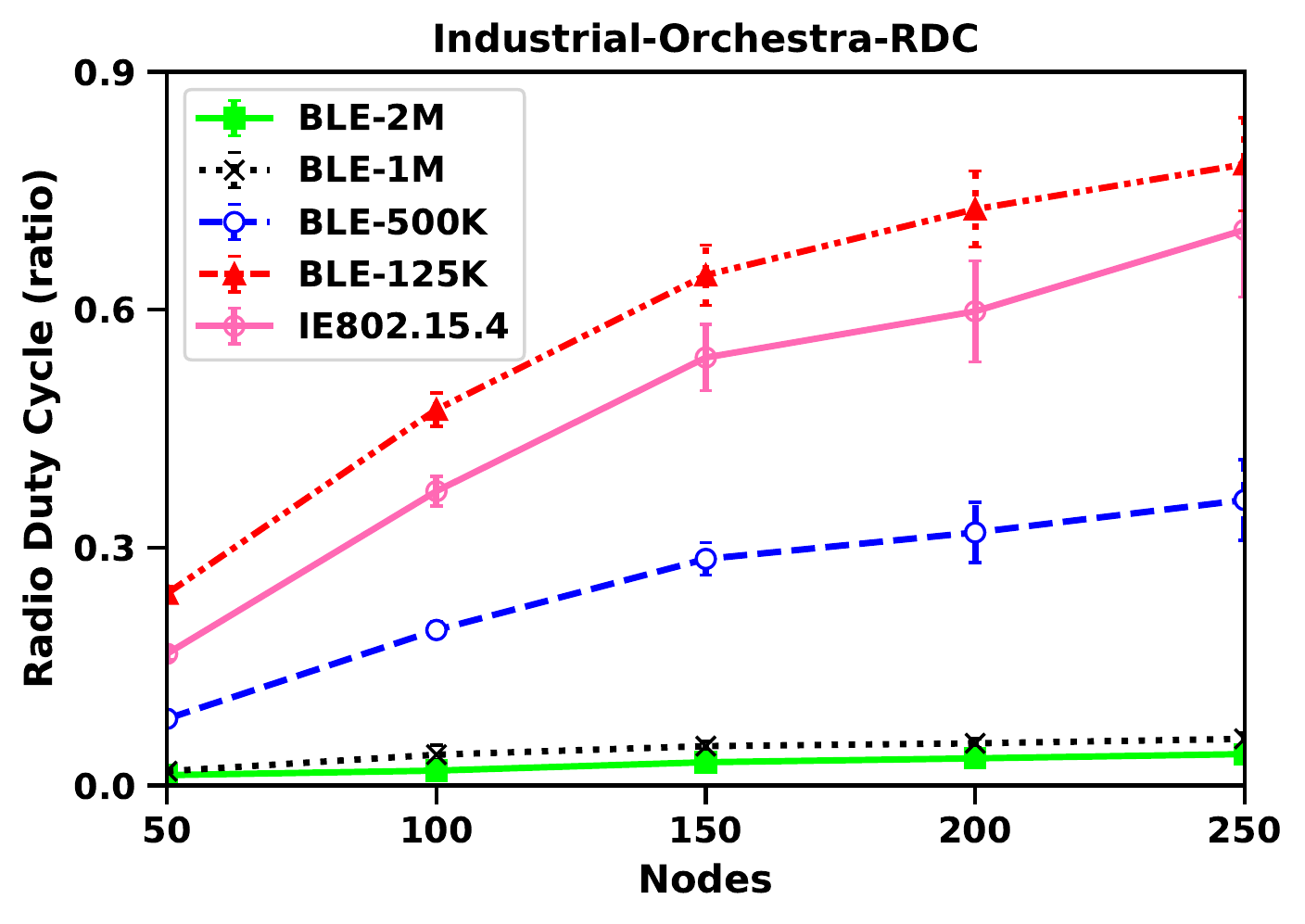}
		\caption{Industrial using Orchestra}
		\label{fig:indu-orch-dc}
	\end{subfigure}%
	\begin{subfigure}[t]{.65\columnwidth}
	    \vspace{+1mm}
		\centering
		\includegraphics[width=1.\columnwidth]{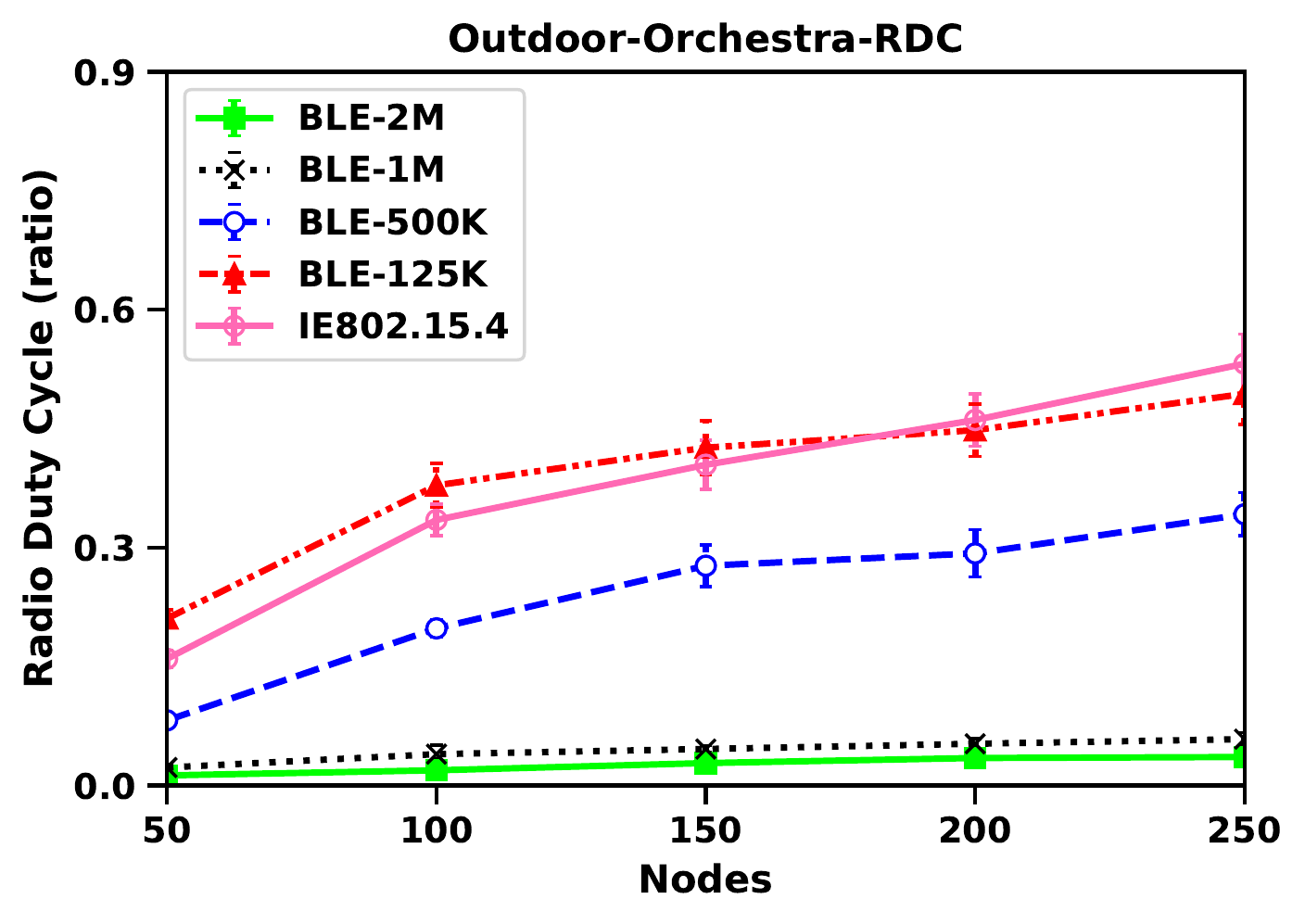}
		\caption{Outdoor using Orchestra}
		\label{fig:out-orch-dc}
	\end{subfigure}%
	\hfill
	\begin{subfigure}[t]{.65\columnwidth}
		\centering
		\includegraphics[width=1.\columnwidth]{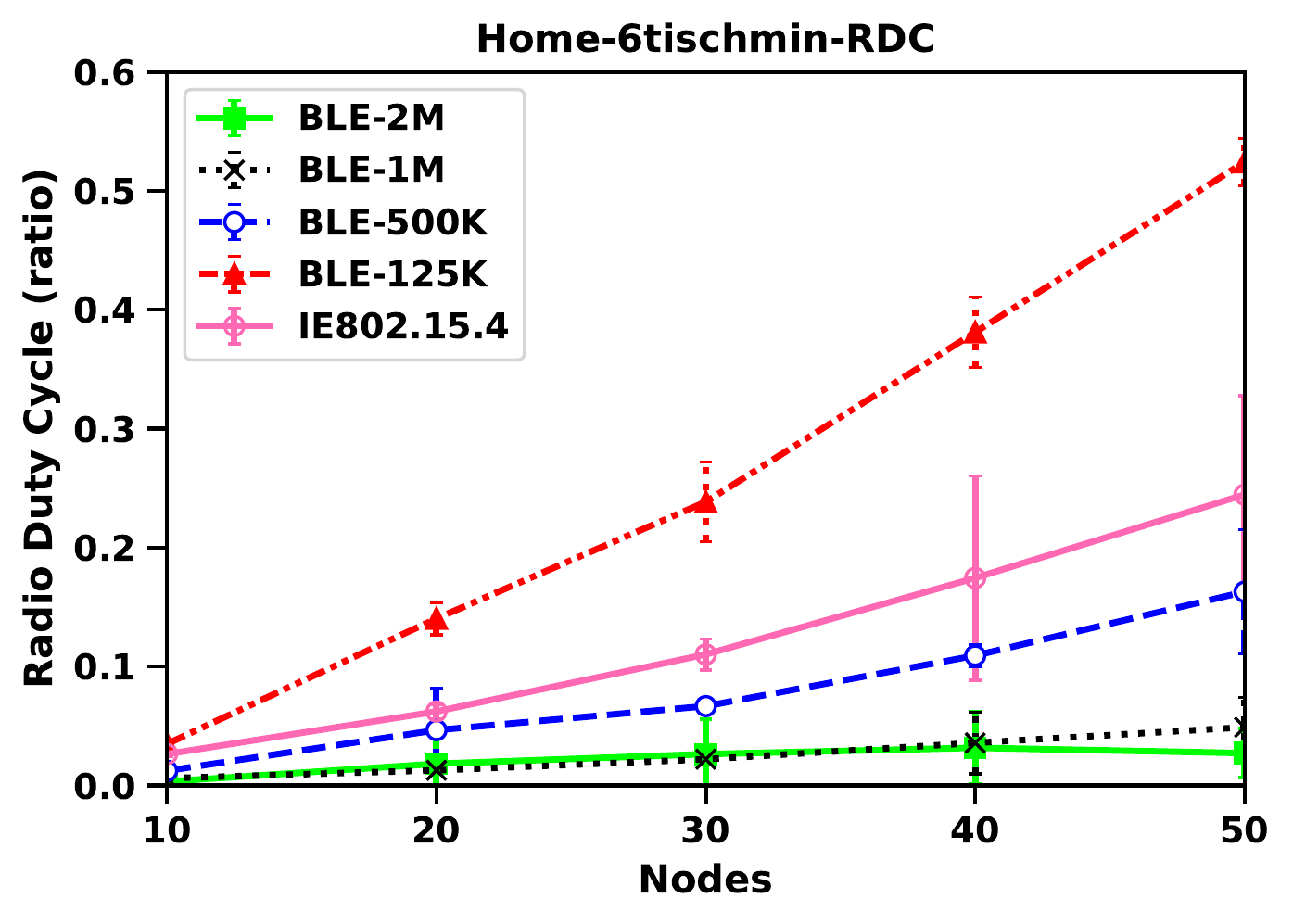}
		\caption{Home using \minimal}
		\label{fig:home-6ti-dc}
	\end{subfigure}
	\begin{subfigure}[t]{.65\columnwidth}
		\centering
		\includegraphics[width=1.\columnwidth]{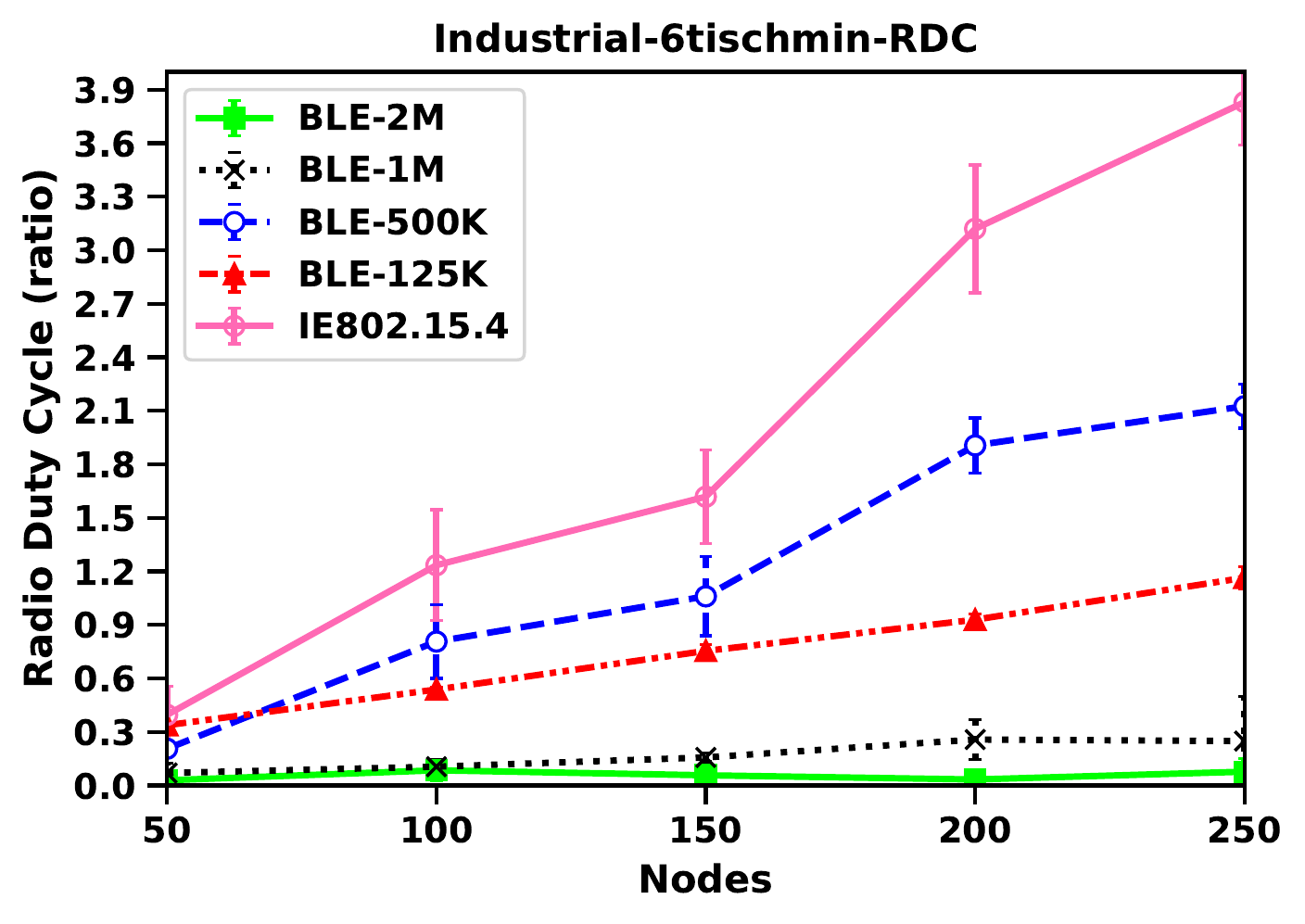}
		\caption{Industrial using \minimal}
		\label{fig:indu-6ti-dc}
	\end{subfigure}
	\begin{subfigure}[t]{.65\columnwidth}
		\centering
		\includegraphics[width=1.\columnwidth]{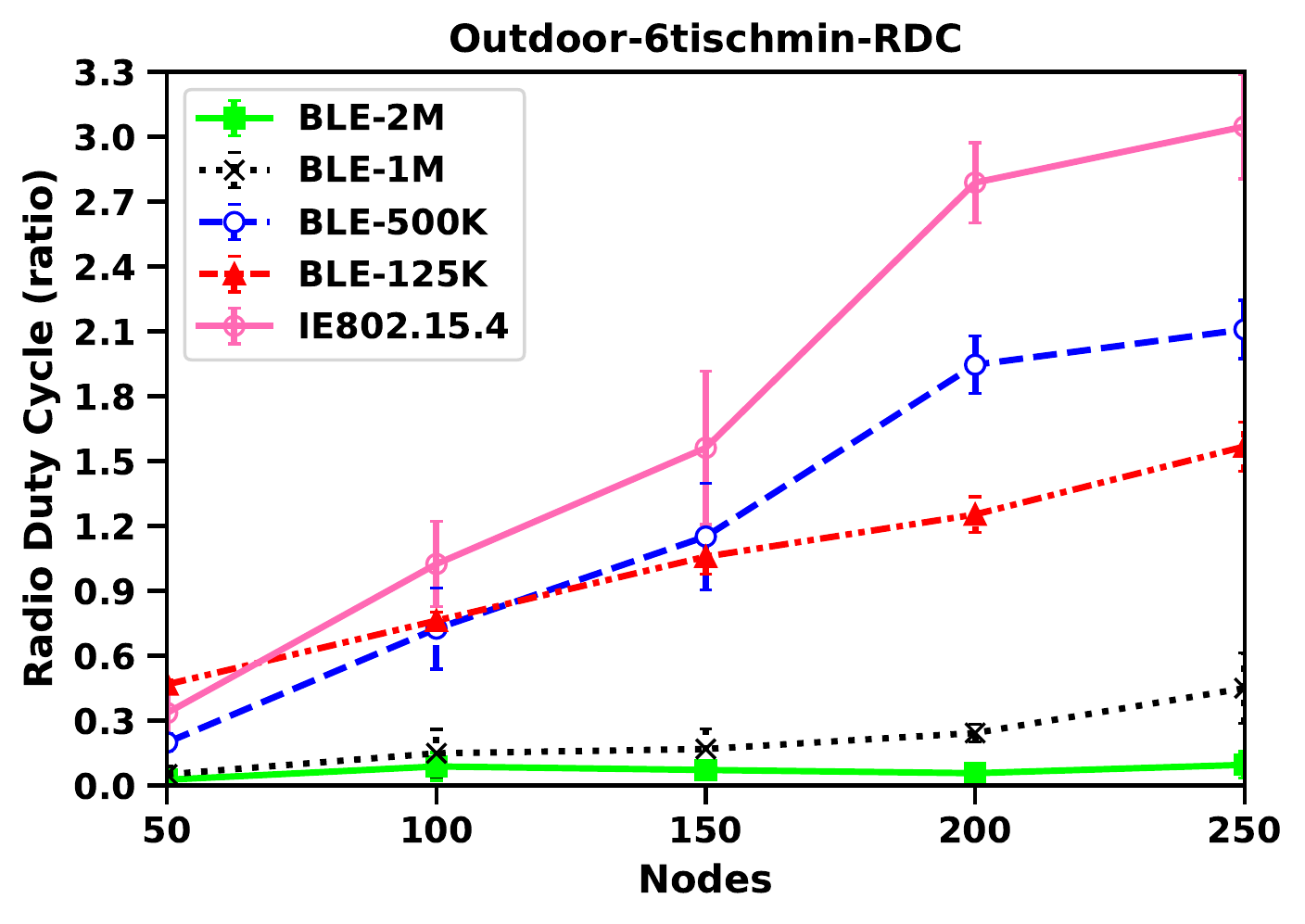}
		\caption{Outdoor using \minimal}
		\label{fig:out-6ti-dc}
	\end{subfigure}
	\vspace{-0.50mm}
	\caption{Radio Duty Cycle.}
	\vspace{-2.0mm}
	\label{fig:dc}
\end{figure*}

\section{Results and Discussion} \label{sec:results}

We discuss the results, making observations with respect to the three RF environments and two 6TiSCH schedulers.

\boldpar{Packet Delivery Ratio (PDR)} Figure~\ref{fig:pdr} shows, in a \textit{home} RF environment, \blefive\,500K has the highest PDR with the Orchestra followed by \ieee, \blefive\,125K, \blefive\,1M, and \blefive\,2M. This trend is similar in \minimal except for the \blefive\,125K. Specifically, Orchestra offers a similar PDR to \ieee, while in \minimal, \blefive\,125K’s PDR drastically decreases as the number of nodes increases. We observe that \blefive\,125K yields the largest collisions in both Orchestra and \minimal with a larger scale in \minimal. A high surge of collisions reflects the packet delivery problems, which explains the lower PDR of \blefive\,125K than expected. Note that \blefive\,125K has the most extended radio range; thus, the highest node degree or neighbors. A transmitting node with such a long-range PHY enables packets to travel fewer hops. However, a large number of links also increases the probability of collisions. 

On the other hand, in the case of \blefive\,2M, the radio range is the shortest one, i.e., low node degree, which seems useful to reduce collisions. However, conversely, \blefive\,2M has the highest transmission rate along with the largest number of hops. In sch a case, packets therefore compete for resources at each forwarding transmission. As the number of nodes increases, the chance of collisions also increases and impacts the PDR. Thus, a PHY that can balance the radio range and packet rates like \blefive\,500K or \ieee becomes the winner.   

In industrial IoT, the PDR trend is similar to that of the home network for the Orchestra scheduler, with a few exceptions. For example, we clearly see two groups of PHYs: \blefive\,500K, \ieee, and \blefive\,125K (is the worst among 3) and \blefive\,1M and 2M. Also, the average PDR is slightly low for all PHYs compared to the home networks. We suspect that it may be due to the industrial environmental settings. Also, the scale of the network in IIoT is way higher than that of the home environment. In the case of \minimal, we observe a similar performance trend like in-home, i.e., \blefive\,125K has the worst performance, whereas \blefive\,500K and \ieee have the best PDR. Finally, when we move to outdoors with the same scale as in industry, we observe a quite similar performance in both the schedulers to that of IIoT. 

Overall, we conclude that Orchestra outperforms \minimal for the \blefive\,125K. Otherwise, their performance is similar in all three environments. In the case of the remaining PHYs, \blefive\,500K or \ieee can safely be deployed to offer high PDR in any of the environments we considered. These two schemes can balance the radio range and traffic rate to reach destinations over a reasonable number of hops with low interference.

\boldpar{Latency} Figure~\ref{fig:latency} presents the average latency in all three environments for the two schedulers. In a home network, while using Orchestra  \blefive\,2M, \blefive\,1M, \blefive\,500K, and \ieee have a very similar latency with \blefive\,2M has a slightly better result, especially with fifty nodes. The latency of \blefive\,125K is significantly higher compared to the rest of the schemes. Also, the variation in latency is quite large. In the \minimal scheduler, the performance trend is similar to that of Orchestra with a lower variation in latency except for \blefive\,125K.  We suspect two factors contribute to this variation: the number of collisions and the backoff algorithm to access the wireless channel. The high number of collisions forces packets to get retransmitted. During that retransmission, packets wait for a random amount of time; once they get the opportunity to grab the channel, there can be collisions again in the presence of high collision probability (e.g., for \blefive\,125K). Also, packets may travel a slightly different route due to the retransmission, impacting the latency.   


Another observation is the impact of number of nodes on the latency, where the performance trend is similar across all environments over the chosen schedulers, i.e., the larger the number of nodes, the higher the latency. For example, the latency increases slightly when the number of nodes increases from ten to fifty. Packets need to traverse a slightly longer route while having a larger network. Also, the probability of collisions increases with the increasing number of nodes and impacts the latency. However, \blefive\,125K is clearly the worst one with an unstable behavior among all PHYs due to its high number of collisions with the longest radio range. Overall, in the home networks, \minimal has a better and stable latency trend (low variation) compared to the Orchestra.

In industrial networks, we clearly identify three groups of PHYs for both the schedulers.  Specifically, \blefive\,2M, \blefive\,1M have the lowest latency (at the ballpark of 500 ms even for a large number of nodes). \blefive\,500K, and \ieee are the next two PHYs with slightly higher latency, while \blefive\,125K has the worst performance. The radio range and hence the number of collisions is the main driving force for such latency trend, i.e., a shorter radio range can avoid collisions significantly and offer a better latency. As home networks, we observe that \minimal has a stable latency trend compared to the Orchestra. However, one interesting point is the performance of \blefive\,125K over \minimal, where the latency drastically degrades for the number of nodes higher than 150. With that high number of nodes, packet delivery rates significantly drop for \blefive\,125K and hence the reported latency from successfully delivered packets. Finally, outdoors offer a performance trend similar to that of the industrial networks.

Overall, the latency trend is similar across different network environments over the Orchestra and \minimal schedulers except for \blefive\,125K. This PHY shows different behavior for the two schedulers. In both cases, the latency is quite high, which we explained above. However, the overall latency trend of \blefive\,125K is different for Orchestra and \minimal. In the former case, the latency increases in a different magnitude with the increasing number of nodes in all three environments. However, in the latter case, the latency starts falling after the number of nodes reaches 100. We suspect that the behavior is related to the high collisions and the low number of successful packet delivery, i.e., the average latency reflects those packets reaching destination perhaps over short routes.

\boldpar{Radio Duty Cycle} Across all environments, the radio duty cycle (RDC) at each PHY increases with respect to the number of nodes (see Figure~ref{fig:dc}). As expected, radio-on time roughly follows the PHY data rates, with the lowest data rate PHYs exhibiting the highest RDC. Interestingly, any difference in RDC is less pronounced on the two \blefive\,1M and 2M PHYs -- despite the latter being twice the rate of the former. Furthermore, there is very little increase in RDC as the network scales, while the reliability results in Figure~\ref{fig:pdr} show that both PHYs suffer significantly in larger networks. This behavior is explained by the fact that the RDC depends on the packet transmission rate. In networks where the majority of packets are lost near the source nodes, the RDC results appear to be better than they would be if all packets would reach their destinations.

In home networks, \blefive\,1M and 2M PHYs offer the lowest RDC while \blefive\,125K has the worst. \blefive\,500K and \ieee in between with the latter one having the worse RDC. An interesting observation is that within the industrial and outdoor environments, the RDC difference between \blefive\,125K and \ieee is far less pronounced over Orchestra. Specifically, in the outdoor environment, \blefive\,125K performs better at extremely large network sizes ($\ge$250 nodes). This trend is likely due to the receiver sensitivity gains on the \blefive\,125K PHY over the DSSS employed in \ieee, which are more advantageous in outdoor LOS scenarios.  Finally, RDC increases significantly for all PHYs when using \minimal compared to Orchestra, at around 1.6x $\sim$ 2.3x. This is to be expected, as Orchestra tries to schedule optimal transmissions based on the RPL DAG, while MSF is analogous to slotted CSMA and uses a shared slot for all traffic.

\section{Conclusions}
\label{sec:conclusions}

We start this work with three hypotheses: (i) that different PHY options are suited for different low-power wireless network applications, (ii) that adapting a network's PHY depending on RF or networking conditions could potentially lead to increased performance than using a single PHY layer, and (iii) that different schedulers may have a different impact on the performance of a PHY .

The results show clear evidence for the first hypothesis: different performance metrics benefit from using different PHY options.
\blefive\,500K is the best option for applications that require high PDR;
in contrast, the uncoded \blefive\,1M and \blefive\,2M options are better for applications that need to be optimized for delay or radio duty cycle. Equally, these two uncoded options greatly reduce the TSCH slot durations, increasing the communication speed and minimizing the radio-on time.
%

The same performance trends appear in \textit{home}, \textit{industrial}, and \textit{outdoor} networks.
We explain the lower-than expected performance of longer-range networks with a high number of collisions.
As the radio range is increased, the average number of neighbors per node increases; consequently, there are more packet collisions and the PDR is reduced. This is especially clear for \blefive\,125K, which has a better PDR when Orchestra is used, as it reduces the number of collisions compared with \minimal. However, the remaining PHYs have a similar performance with both the schedulers.

Finally, to the best of our knowledge, this work also is the first to directly compare, in identical settings and at a scale of 100s of nodes, 
\blefive-based 6TiSCH networks all four \blefive PHY options with \ieee in multihop low-power wireless 6TiSCH networks. The results show that \blefive\,500K shows consistently higher PDR and lower energy consumption than \ieee in nearly all of our experiments, while keeping the delay similarly low. Based on these discoveries, we plan to design a high performance multi-PHY 6TiSCH IoT network protocol in our future work.

\bibliographystyle{IEEEtranN}
\bibliography{references}
\end{document}